\documentclass[letterpaper]{article}

\usepackage[header]{aimatters}
\usepackage{natbib}
\usepackage{color}
\usepackage{verbatim}

\title{Mechanism Design for Social Good}
\author{
\AIMauthor{Rediet Abebe}{Cornell University}{red@cs.cornell.edu}
\AIMauthor{Kira Goldner}{University of Washington}{kgoldner@cs.washington.edu}
}

\AIMbio{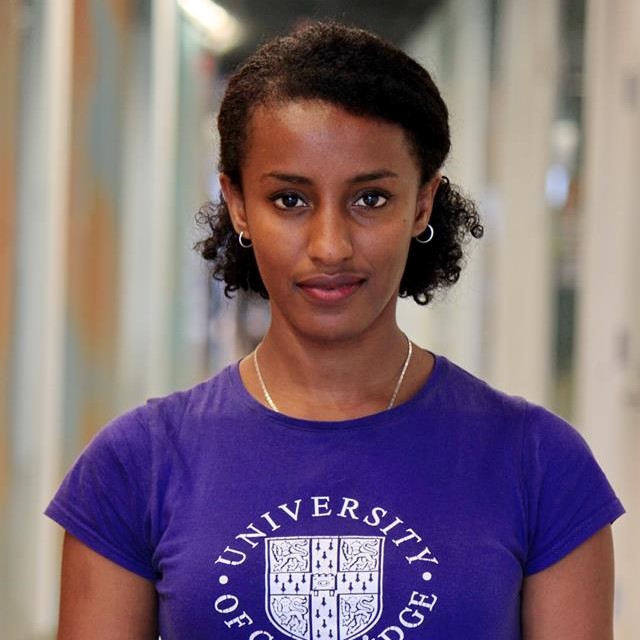}{\textbf{Rediet Abebe} is a PhD candidate in computer science, advised by Jon Kleinberg. Her research focuses on algorithms, artificial intelligence, and their applications to social good. In particular, she uses algorithmic and computational insights to better understand socioeconomic inequality and inform interventions for improving access to opportunity. Her work is generously supported by fellowships and scholarships through Facebook, Google, and the Cornell Graduate School. Prior to Cornell, she completed an M.S. and a B.A. in Applied Mathematics and Mathematics from Harvard University and an M.A. in Mathematics from the University of Cambridge. She was born and raised in Addis Ababa, Ethiopia.}

\AIMbio{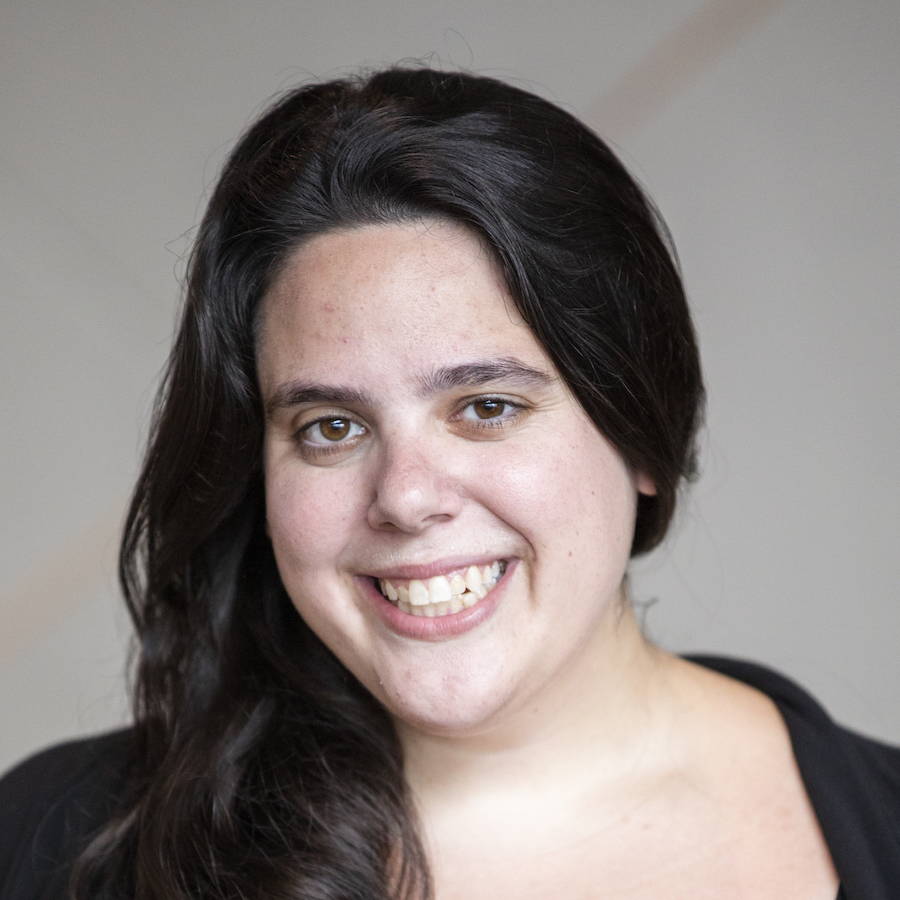}{\textbf{Kira Goldner} is a fifth-year PhD student at the University of Washington in the Department of Computer Science and Engineering, advised by Anna Karlin. Her research focuses on problems in mechanism design, particularly on (1) maximizing revenue in settings that are motivated by practice and (2) on mechanism design within health insurance and online labor markets. She is a 2017-19 recipient of the Microsoft Research PhD Fellowship and was a 2016 recipient of a Google Anita Borg Scholarship. Kira received her B.A. in Mathematics from Oberlin College and also studied at Budapest Semesters in Mathematics.}

\begin{document}
\maketitle

\section{Introduction}

Across various domains---such as health, education, and housing---improving societal welfare involves allocating resources, setting policies, targeting interventions, and regulating activities. These solutions have an immense impact on the day-to-day lives of individuals, whether in the form of access to quality health-care, labor market outcomes, or how votes are accounted for in a democratic society. Problems that can have an outsized impact on individuals whose opportunities have historically been limited often pose conceptual and technical challenges, requiring insights from many disciplines. Conversely, the lack of interdisciplinary approach can leave these urgent needs unaddressed and can even exacerbate underlying socioeconomic inequalities.

To realize the opportunities in these domains, we need to correctly set objectives and reason about human behavior and actions. Doing so requires a deep grounding in the field of interest and collaboration with domain experts who understand the societal implications and feasibility of proposed solutions. These insights can play an instrumental role in proposing algorithmically-informed policies. In many cases, the input data for our algorithms may be generated by strategic and self-interested individuals who have a stake in the outcome of the algorithm. To get around this issue, we can deploy techniques from \emph{mechanism design}, which uses game theory to align incentives or analyze the strategic behavior of individuals who interact with the algorithms. 

The \emph{Mechanism Design for Social Good} (MD4SG) research agenda is to address problems for which insights from algorithms, optimization, and mechanism design have the potential to improve access to opportunity. These include allocating affordable housing services, designing efficient health insurance markets, setting subsidies to alleviate economic inequality, and several other issues affecting many individuals' livelihoods. This research area falls at the interface of artificial intelligence, theoretical computer science, and the social sciences. Since the fall of 2016, the authors of this piece have been co-organizing the Mechanism Design for Social Good research group, workshop series, and colloquium series \cite{md4sg,md4sgworkshop}. The group comprises a large network of researchers from various disciplines, including computer science, economics, sociology, operations research, and public policy. Members of the group partner with domain experts in non-government organizations, think tanks, companies, and other entities with a shared mission. The mission is to explore new frontiers, garner interest in directions in which algorithmic and mechanism design insights have been under-utilized but have the potential to inform innovative interventions, and highlight exemplary work.

In this piece, we discuss three exciting research avenues within MD4SG. For each of these, we showcase ongoing work, underline new directions, and discuss potential for implementing existing work in practice.

\section{Access to Opportunity in the Developing World}

New technologies and data sources are frequently leveraged to understand, evaluate, and address societal concerns across the world. In many developing nations, however, there is a lack of information regarding underlying matters---whether that be the prevalence of diseases or accurate measurements of economic welfare and poverty---due to the unavailability of high-quality, comprehensive, and reliable data \cite{UN}. This limits the implementation of effective policies and interventions. An emerging solution, which has been successfully demonstrated by the Information Communication Technology for Development (ICT4D) research community, has been to take advantage of high phone and Internet penetration rates across developing nations to design new technologies which enable collection and sharing of high-quality data. There has also been recent work from within the AI community to use new data sources to close this information gap \cite{africasearch,povertymapping}. Such AI-driven approaches surface new algorithmic, modeling, and mechanism design questions to improve the lives of many under-served individuals. 

A prominent example is in agriculture, which accounts for a large portion of the economy in many developing nations. Here, viral disease attacks on crops is a leading cause of food insecurity and poverty. Traditional disease surveillance methods fail to provide adequate information to curtail the impact of diseases \cite{mwebaze3,mwebaze2,mwebaze1}. The Cassava Adhoc Surveillance Project from Makerere University implements crowd-sourcing surveillance using pictures taken by mobile phones in order to address this gap \cite{mutembesa}. The tool is set up as a game between farmers and other collaborators, and aims to collect truthful, high-value data (e.g., data from hard-to-reach locations). This approach underlines interesting challenges, such as how to optimally incentivize individuals to collect high-quality information and how to augment this information with existing methods. Similar issues arise in other domains---e.g., in citizen science and in computational sustainability \cite{birds1,birds2}. Finding solutions in the context of the developing world may therefore have a broader global impact. 
 
Lack of information also leads to inefficiencies in existing systems, presenting a possibility to introduce solutions that abide  by existing cultural and technological constraints. For instance, large price discrepancies and major arbitrage opportunities present in markets for agricultural products in Uganda suggest large market inefficiencies \cite{kudu1}. To alleviate this, \cite{kudu2} introduce Kudu---a mobile technology that functions over feature phones via SMS service. Kudu facilitates transactions between farmers in rural areas and buyers at markets in cities by allowing sellers and buyers to post their asks and offers. Kudu has been adapted by users across Uganda and many trades have been realized through this system. 

Availability of new technologies also presents opportunities to tackle fundamental problems related to poverty. Advances in last-mile payment technologies, for example, enable large-scale, secure cash transfers. GiveDirectly leverages this and the popularity of mobile money across the world to create a system where donors can directly transfer cash to recipients \cite{GD,gd2}. GiveDirectly moves the decision about how to use aid from policy-makers to recipients, giving recipients maximum flexibility. Such aid generates heterogeneity in outcomes---e.g., families may use aid to start a business, pay rent, cover health costs, and so on. Policy-makers used to prioritizing specific outcomes may be uncomfortable by such a model. A research question then is: can we predict how a given population will use aid? Likewise, how can we target people for whom the interventions will make the largest difference? Aid has historically been targeted on the basis of finding the most deprived people. The ability to model heterogeneous treatment effects opens the door for designing more nuanced mechanisms that fairly and efficiently allocate subsidies in order to maximize a desired outcome. 

Problems in the developing world surface unique challenges at the intersection of AI, ICT4D, and development economics. Solutions often have to be implemented in resource-constrained environments (e.g., over feature phones or with low network connectivity) \cite{nicki,tapan}. Key populations of interest (e.g., women, people living in rural parts, individuals with disabilities) may not be easily accessible \cite{women,blind,rural}. Individuals may have low-literacy \cite{literacy}. Lack of understanding of socio-cultural norms and politics, furthermore, may inhibit proposed interventions \cite{aditya}. All of these highlight the need for a multi-stakeholder approach that leverages technological advances, innovative technical solutions, and partnerships with individuals and organizations that will be impacted by the solutions. MD4SG fosters one such environment in which insights from across these disciplines inform the design of algorithms and mechanisms to improve the lives of individuals across the world.

\section{Labor, Platforms, and Discrimination}

Online platforms are ubiquitous, providing a vast playground for algorithm design and artificial intelligence. Every policy decision, however, impacts and interacts with the platform's strategic users. In this section, we will focus on online labor markets and how discrimination effects stem from a platform's decisions. Past work begins to investigate some aspects of platforms, of strategic agents, and of discrimination in labor markets, but there are still major opportunities for work at the intersection, and insights from mechanism design and AI are ripe for the job.

One central issue surrounding labor markets is that of \emph{hiring}, in which a firm takes information about a potential candidate and makes an employment decision. Firms act as classifiers, labeling each applicant as ``hire" or ``not hire" based on an applicant's ``features," such as educational investment or a worker's productivity reputation. In the process of making hiring decisions, however, the firm may potentially make discriminatory decisions---perhaps by using protected attributes, or by not correcting for differences in applications that stem from systemic discrimination \cite{sendhil,genderbias}. Bias in hiring decisions may arise due to implicit human bias or algorithmic bias, in which algorithms replicate human and/or historic discrimination that is reflected in the data on which they are trained \cite{au,ai,ao,wmd}. 

One recent line of work investigates hiring policies that achieve diversity or statistical parity (with respect to certain groups) among the hired workers, and how workers make their investment decisions (e.g. whether to attend college) based on the hiring policies in place. \cite{coateloury,loury,lily} study settings where there is some known underlying bias or historical discrimination against certain groups; the aim is to characterize hiring policies that are optimal-subject-to-fair-hiring, and to quantify any loss in efficiency compared to optimal-but-discriminatory policies. These works explore two settings: first, when hiring decisions must be ``group-blind," that is, they cannot take group membership into account, and second, when they are ``group-aware". The aim is to choose hiring policies that will mitigate discrimination against protected categories. \cite{lily} highlights additional complexity that arises in dynamic settings where workers are hired based on investment decisions (e.g. college GPA) in an initial temporary labor market (e.g. internships) and this job creates a worker's initial productivity reputation that is then used in the permanent labor market. Many of these findings also discuss ``trade-offs" between group-blind and group-aware policies.

Another aspect of labor markets is that a worker may have the ability to pay to change a feature of her application in some illegitimate or unfair way in order to improve her outcome in the labor market. 
\cite{hardt} examine this problem from a robust machine learning perspective.  
Under certain assumptions of the cost required to change an applicant's reputation, they characterize classifiers that optimally compare to the original reputation (before the applicant modified it).

These are only two aspects at the interplay between hiring and strategic agents; hiring, furthermore, is only one aspect of the labor market. Consider today's popular online labor markets, such as Mechanical Turk, Upwork, Task Rabbit, and Lyft, in which the platform's goal is to match workers to employers or jobs. In these labor markets, the platform's decisions, even at a granular level, have a large impact on the workers and firms. Consider the following platform decisions. Visibility: How many firms can workers see at a time? What capacity do they have to search job offers? Can workers see jobs and jobs see workers? Initiation: Which side (or both) can submit applications? Initiate messaging? Set contract terms? Information: What information is displayed about parties on the opposite side? Name? Photo? Ethnicity? Wage history? Reputation? 

Each of these decisions impacts the outcome---not only the quality of the match, but also whether (and how much) discrimination occurs.  In a recent paper, \cite{karensolon} outline categories of platform decisions which may mitigate or perpetuate discrimination in labor markets, including the high-level categories of setting platform discrimination policies or norms, structuring information and interactions, and monitoring/evaluating discriminatory conduct.  

In offline labor markets, it may be challenging or infeasible to collect data to understand the nature and extent of discrimination. Online labor markets, on the other hand, yield rich data about employer-employee interactions and present the possibility of conducting experiments aimed at reducing bias and discrimination or other desired societal objectives. For instance, \cite{hortonwagehistory} look at the impact on hiring of hiding workers' wage history.  \cite{hortonelicitation,hortonworkercap} look at the impact of trying to elicit additional information (features) from workers or firms, and the impact of this strategically-reported information on hiring.  \cite{hortonminwage} examines who the hired worker population is when a minimum wage is imposed on one platform. Each of these provide insights into labor dynamics that may inform platform design and interventions. 

Online labor markets provide a rich playground for techniques from algorithms, AI, and mechanism design to study how each aspect of platform design impacts discriminatory effects, workers' actions, and the desired objective for the platform.

\section{Allocating Housing and Homelessness Resources}

Allocation of resources---such as public housing, housing vouchers, and homelessness services---has a long history in the economics and computation literature. Even simple-to-state problems here have given rise to challenging research questions, many of which are still open. Increased scarcity of housing resources, growing need for services, and the use of algorithmic decision-making tools all open up several avenues with major opportunities for reforming policies and regulations. Here, we discuss some foundational work, new challenges, and opportunities that emerge at the nexus of algorithm and mechanism design, AI, and the social sciences. 

Millions of individuals across the US have been evicted or are at risk of experiencing eviction every year. In groundbreaking work, \cite{desmond2,desmond} shows that eviction is much more common than was previously documented. By compiling the first ever evictions database, Desmond shows that there is an estimate of 2.3 million evictions in 2016 alone and argues that eviction is a cause of poverty \cite{EL}. Using this database, and other similar datasets, we may be able to employ a combination of machine learning and statistical techniques to gain a better understanding of what causes housing instability and homelessness. We may then be able to build on this work to design algorithms and mechanisms that can improve on allocation policies. For instance, \cite{sanmay} uses counter-factual predictions to improve homelessness service provisions. By doing so, they realize some gains on reducing the number of families experiencing repeated episodes of homelessness. At the same time, \cite{ai} emphasizes that caution must be taken when using automated decision-making tools for allocating limited resources in such high-stakes scenarios. Eubanks argues that such tools may be used to reduce failure rates by caseworkers; but, if not approached with care, they can deepen already existing inequalities. Furthermore, such tools alone are limited: they do not address the lack of housing and homelessness resources or eliminate human biases or discrimination.  It is therefore crucial to take advantage of the confluence of insights from cross many disciplines in order to serve the needs of such vulnerable populations. 

An issue that is growing in prominence in housing contexts is that of information. Little is documented about how landlords or housing authorities screen applications and make decisions. \cite{ambrose} show that there is increased restriction in access to rental housing since landlords mitigate information asymmetry by investing in screening tenants. With the increased use and availability of data about individuals, it is of paramount importance to understand the role of information in the decision-making process of entities, such as landlords or housing agencies, who have enormous discretion in how and whether families are housed. 

Although the introduction of automated tools introduces acute challenges related to housing, the use of algorithmic techniques dates back several decades and there are many fundamental problems that remain unsolved. An early work here is that of \cite{hz}, which considers the ``house allocation problem" of assigning each individual to one item, such as a house. They introduce a mechanism which satisfies natural efficiency and fairness notions but is not incentive-compatible. That is, individuals may be able to improve their outcome by misreporting their true preferences. Since then, several mechanisms have been proposed, including the popular Randomized Serial Dictatorship (RSD) mechanism, which uses a random lottery \cite{rsd}. This mechanism is incentive-compatible, but fails to satisfy the fairness criteria of \cite{hz}. It is used as a standard mechanism in many domains, including housing. An outstanding question related to these is then the design of incentive-compatible, fair, and efficient mechanisms for the house allocation problem. 

Due to increased scarcity of resources, allocation protocols often involves waiting lists and priority groups. Policy constraints make wait-list design a dynamic rationing problem rather than the static assignment problem discussed above. Dynamic mechanisms present several technical and practical challenges; e.g., incentive-compatibility may be infeasible in dynamic settings due to waiting time trade-offs for applicants. There are consequential design decisions related to how to manage waitlists and different metropolitan areas have different policies (e.g., setting priority groups, conditions under which individuals are removed from the waiting list, set of choices, and many others). Each of these policies impacts the dynamics of the allocation process, waiting time, and quality of matches. Recent work has studied how to design mechanisms satisfying various desiderata and quantify differences in quality of matches across various mechanisms \cite{peng,leshno,neil,dan}.

\section{Conclusion}

As the use of algorithmic and AI techniques becomes more pervasive, there is a growing appreciation of the fact that the most impactful solutions often fall at the interface of various disciplines. The Mechanism Design for Social Good research agenda is to foster an environment in which insights from algorithms and mechanism design can, in conjunction with the social sciences, be used to improve access to opportunity, especially for communities of individuals for whom opportunities have historically been limited. In this piece, we have highlighted MD4SG research avenues related to issues in developing nations, labor markets, and housing. For each of these, we have discussed the need to work in close partnership with a wide range of stakeholders to set objectives that best address the needs of individuals and propose feasible solutions with desired societal outcomes. There are numerous other domains in which this kind of interdisciplinary approach for designing algorithms and mechanisms can improve the lives of many individuals; we invite readers to learn more through our colloquium and workshop series.

\section{Acknowledgments}
We are indebted to members of the Mechanism Design for Social Good research group--Ellora Derenoncourt, Alon Eden, Lily Hu, Manish Raghavan, Sam Taggart, Daniel Waldinger, and Matt Weinberg---our co-organizer Irene Lo, and our advisors Anna Karlin and Jon Kleinberg for their generosity in sharing their knowledge and support throughout the past two years. We additionally thank Mutembesa Daniel, Paul Niehaus, Fabian Okeke, and Aditya Vashistha for helpful discussions and pointers. We are grateful, as ever, for the numerous researchers in the economics and computation research communities for their enthusiastic support and guidance throughout the development of the group and workshop series.

\bibliographystyle{plainnat}
\bibliography{refs}

\end{document}